\documentclass[prd,aps,floats,amssymb, secnumarabic ,nofootinbib,preprintnumbers]{revtex4}
\usepackage{natbib}
\usepackage{dcolumn}
\usepackage{graphicx}

\newcommand{\be}{\begin{equation}}
\newcommand{\ee}{\end{equation}}
\newcommand{\bea}{\begin{eqnarray}}
\newcommand{\eea}{\end{eqnarray}}
\begin{document}
%
%
\title{Supernova, baryon acoustic oscillations, and CMB surface distance constraints on $f(G)$ higher order gravity models}
\author{Jacob Moldenhauer$^1$\footnote{Electronic address: jam042100@utdallas.edu}, Mustapha Ishak$^1$\footnote{Electronic address: mishak@utdallas.edu}, John Thompson$^1$, Damien A. Easson$^{2,3}$\footnote{Electronic address: easson@asu.edu}}
\affiliation{
$^1$Department of Physics, The University of Texas at Dallas,\\ Richardson, TX 75083, USA\\
$^2$Department of Physics  \& School of Earth and Space Exploration  \& Beyond Center,
Arizona State University, Tempe, AZ, 85287-1404, USA\\
$^3$Institute for the Physics and Mathematics 
of the Universe, University of Tokyo,
5-1-5 Kashiwanoha, Kashiwa, Chiba 277-8568, Japan}
\date{\today}
\begin{abstract}
We consider recently proposed higher order gravity models where the action is built from the Einstein-Hilbert action plus a function $f(G)$ of the Gauss-Bonnet invariant. The models were previously shown to pass physical acceptability conditions as well as solar system tests. In this paper, we compare the models to combined data sets of supernovae, baryon acoustic oscillations, and constraints from the CMB surface of last scattering. 
We find that the models provide fits to the data that are close to those of the LCDM concordance model. The results provide a pool of higher order gravity models that pass these tests and need to be compared to constraints from large scale structure and full CMB analysis.  
\end{abstract}
\maketitle
%
\section{introduction}
%
Higher order gravity models have been proposed among possible causes of late-time cosmic acceleration \cite{HOGpapers}. These models are built out of higher order curvature invariants that yield generalized field equations with a coupling between the mass content of the universe and the space-time curvature that produces a late-time self accelerating phase.  
A large body of papers have been devoted to the so-called $f(R)$ \cite{FRpapers} models while a smaller fraction study models based on invariants built out of the Ricci and Riemann tensors \cite{RicciRiemannpapers}.

In addition to the phenomenology of an accelerating cosmic expansion, higher-order gravity models have theoretical motivations within unification theories of fundamental interactions, and through field quantization on curved space-times \cite{BirrellDavies1982, Weinberg1995}. Indeed, higher-order invariants appear automatically in most quantum gravity proposals \cite{DeWitt1965, Ashtekar1981, Weinberg1995}, string theories \cite{Polshinski19982000, GreenSchwarzWitten19871999}, supergravity \cite{Brandt2002, ChamseddineArnowittNath1982}, and loop quantum gravity/cosmology \cite{AshtekarLewandowski2004, Rovelli1998, DateSengupta2009}. In these theories, high-order loop corrections on curved space-time are related to higher-order combinations of the Riemann curvature invariants. Higher-order invariants have been actively discussed in relation to avoiding cosmological curvature singularities (see \cite{SingularityPapers} and references therein). In some of these theories, the invariants are regrouped in a topological invariant combination called the Gauss-Bonnet invariant, denoted as $G$. This combination gives a theory free of unphysical states \cite{DeWitt1965, Zwiebach1985, LiBarrowMota2007}. 

In this paper, we study cosmological constraints on some models where the action is made of the Einstein-Hilbert action plus a function $f(G)$ of the Gauss-Bonnet invariant. We focus on models that have been shown in previous literature to be free of ghost instabilities and superluminal propagations \cite{GBpapers} in cosmological homogeneous and isotropic backgrounds. We perform coordinate transformations in order to write the dynamical equations in a form integrable using numerical schemes and then compare the models to recent observations of supernova magnitude-redshift data, distance to the CMB surface, and Baryon Acoustic Oscillations (BAO), including Hubble Key project and age constraints. 
%
\section{$f(G)$ models} 
%
The models that we investigate in this paper are derived from varying the action 
\be
I=\int{d^{4}x \sqrt{-g}\left[\frac{1}{2}R+f(G)\right]+\int{d^{4}x}\sqrt{-g}L_m}+\int{d^{4}x}\sqrt{-g}L_{rad}
\label{eq:ActionGB}
\ee
with respect to the metric $\delta g_{\alpha\beta}$, where 
\be
G=R^2-4R^{\alpha\beta}R_{\alpha\beta}+R^{\alpha\beta\gamma\delta}R_{\alpha\beta\gamma\delta}
\label{eq:GB}
\ee
is the Gauss-Bonnet invariant, $R$ is the Ricci scalar, $R_{\alpha\beta}$ is the Ricci tensor, $R_{\alpha\beta\gamma\delta}$ is the Riemann tensor, $L_m$ and $L_{rad}$ are the matter and radiation energy Lagrangians, respectively. We 
will work in units with reduced Planck mass $M_{pl}^2 = (8 \pi G_N)^{-1}=1$. The corresponding field equations read
\bea
&&{8[R_{\alpha\gamma\beta\delta}+R_{\gamma\beta}g_{\delta\alpha}-R_{\gamma\delta}g_{\beta\alpha}-R_{\alpha\beta}g_{\delta\gamma}+R_{\alpha\delta}g_{\beta\gamma} +\frac{1}{2}R(g_{\alpha\beta}g_{\delta\gamma}-g_{\alpha\delta}g_{\beta\gamma})]\nabla^{\gamma}\nabla^{\delta}}f_G\nonumber\\
&&+\,(Gf_G-f)g_{\alpha\beta} + R_{\alpha\beta}-\frac{1}{2}g_{\alpha\beta}R= T_{\alpha\beta},
\label{eq:FieldEquationsGB}
\eea
where we use the definition $f_{G}\equiv\frac{\partial{f}}{\partial{G}}$.

Now using the Friedmann-Lemaitre-Robertson-Walker flat metric 
\be
ds^2=-dt^2+a(t)^2d\vec{x}^2
\ee
and Universe filled with matter and radiation, one derives the generalized Friedmann equation
\be
3H^2=Gf_{G}-f{G}-24H^3\dot{f}_{G}+\rho_{m} +\rho_{rad}.
\label{eq:FriedmannGB1}
\ee
where $\rho_m$ and $\rho_{rad}$ are the matter and radiation energy densities, respectively, $H=\dot{a}/a$ and a dot represents $d/dt$. We also note that in terms of $H$, 
\be
R=6(\dot{H}+2H^2)
\ee
and
\be
G=24H^2(\dot{H}+H^2).
\ee
We explore explicit models in the next sections. 
%
%
\section{Recently proposed viable $f(G)$ models}
%
%
\subsection{$f(G)$ models proposed by De Felice and Tsujikawa}
%
In Ref. \cite{DeFelice2008}, the authors imposed certain conditions on the function $f(G)$ and its derivatives. The most important condition being $d^2f/dG^2>0$ in order to ensure the stability of a late-time de Sitter solution as well as the existence of a radiation/matter dominated epochs preceding a late-time accelerating phase. Other additional regularity and viability conditions in \cite{DeFelice2008} single out the following $f(G)$ functions  
\bea
\text{Model DFT-A:}&& f(G)=\lambda \frac{G}{\sqrt{G_0}}\arctan\Big(\frac{G}{G_0}\Big)-\frac{\lambda}{2}\sqrt{G_0}\ln{\Big(1+\frac{G^2}{G_0^2}\Big)}-\alpha \lambda \sqrt{G_0}, 
\label{eq:DeFelicemodelsA}
\\
\text{Model DFT-B:}&& f(G)=\lambda \frac{G}{\sqrt{G_0}}\arctan\Big(\frac{G}{G_0}\Big)-\alpha \lambda \sqrt{G_0},
\label{eq:DeFelicemodelsB}
\\
\text{Model DFT-C:}&& f(G)=\lambda \sqrt{G} \ln{\Big[\cosh\Big(\frac{G}{G_0}\Big)\Big]}-\alpha \lambda \sqrt{G_0},
\label{eq:DeFelicemodelsC}
\eea
where $\lambda$, $G_0$ and $n$ are positive constants and $\alpha$ is a constant. These functions were derived from the integration of the following second order derivatives that satisfy the condition $f_{,GG}>0$ for all values of $G$ along with other regularity conditions \cite{DeFelice2008}.  
\bea
\text{Model DFT-A:}&& f_{,GG}=\frac{\lambda}{G_0^{3/2}[1+(G^2/G_0^2)^n]},\,\, n=1
\label{eq:DeFelice2ndDerivativesGA}
\\
\text{Model DFT-B:}&& f_{,GG}=\frac{2\lambda}{G_0^{3/2}[1+(G^2/G_0^2)^n]},\,\, n=2
\label{eq:DeFelice2ndDerivativesGB}
\\
\text{Model DFT-C:}&& f_{,GG}=\frac{\lambda}{G_0^{3/2}}[1-\tanh^2(G/G_0)], 
\label{eq:DeFelice2ndDerivativesGC}
\eea
Varying the corresponding actions with respect to the metric, the generalized Friedmann equations follow as:

\noindent Model DFT-A:
\bea
\lefteqn{3H^2 = -\sqrt{G_0}\Big(1152H^6(H(4H'+3H+H'')-\alpha(H+H')^2)}\nonumber\\ 
& &-576H^6\ln{\Big(1+\frac{G^2}{G_0^2}\Big)}(H+H')^2- \ln{\Big(1+\frac{G^2}{G_0^2}\Big)}G_0^2-2\alpha G_0^2\Big)\lambda\Big/
 2(G_0^2+G^2)\nonumber\\
&&+\frac{3H_0^2\Omega_m}{e^{3N}} + \frac{3H_0^2\Omega_{rad}}{e^{4N}}+\Lambda,
\label{eq:DeFeliceFreidmannA}
\eea
Model DFT-B:
\bea
\lefteqn{3H^2 = \Big(\alpha G_0^4-576G_0^2H^6(H'+H)(5H'+H)+1152\alpha G_0^2H^6(H'+H)^2}\nonumber\\
& &+331776H^{12}(H'+H)^4(1+\alpha)\Big)\sqrt{G_0}\lambda\Big/ (G_0^2+G^2)^2+\frac{3H_0^2\Omega_m}{e^{3N}} + \frac{3H_0^2\Omega_{rad}}{e^{4N}}+\Lambda,
\label{eq:DeFeliceFreidmannB}
\eea
Model DFT-C:
\bea
\lefteqn{3H^2=\lambda\Big(24\sinh\Big(\frac{G}{G_0}\Big)G_0H^3\cosh\Big(\frac{G}{G_0}\Big)(H'+H)}\nonumber\\
& &-576H^6(3H'^2-H(4H'+H''))\nonumber\\
& & +G_0^2\cosh\Big(\frac{G}{G_0}\Big)^2\Big(\alpha-\ln\Big(\cosh\Big(\frac{G}{G_0}\Big)\Big)\Big)\Big)\Big/G_0^{(3/2)}\cosh\Big(\frac{G}{G_0}\Big)^2\nonumber\\
& &+\frac{3H_0^2\Omega_m}{e^{3N}} + \frac{3H_0^2\Omega_{rad}}{e^{4N}}+\Lambda,
\label{eq:DeFeliceFreidmannC}
\eea
where $\Lambda=\alpha \lambda \sqrt{G_0}$ and we define $N=\ln{a}$ with $'=d/dN$.  As discussed in \cite{Mena2006,MoldenhauerIshak2009}, these higher order gravity models have generalized Friedmann equations that are very stiff ordinary differential equations (ODEs) that require us to use well-adapted numerical codes and logarithmic variables. Thus, we replaced $H$ in the ODEs by the logarithmic variable $u=\ln{(H/\hat{\mu})}$ where $\hat{\mu}$ is a constant normalizing the Hubble parameter. This allows one to write the source terms, see \cite{Mena2006}, as  
\be
\tilde{u}=\ln{(\tilde{\omega}_{rad} e^{-4N} + \tilde{\omega}_m e^{-3N})}/2
\ee
 and 
\be
\tilde{\omega}_m\equiv \frac{8\pi G}{3}\frac{\rho_0}{\hat{\mu}^2}.
\label{eq:wbarsource}
\ee
Further, with $\omega_m = \Omega_m h^2$ and $h=H_0/(100 km/s/Mpc)$, one writes 
\be
\tilde{\omega}_m=\frac{\omega_m}{\hat{\mu}^2}.
\label{eq:wbar}
\ee
Similarly, $\tilde{\omega}_{rad}$ is defined for radiation but we consider its contribution to be negligible at late times.

In terms of the logarithmic variables, the generalized Friedmann equations (equations (\ref{eq:DeFeliceFreidmannA}) (\ref{eq:DeFeliceFreidmannB}) (\ref{eq:DeFeliceFreidmannC})) may be written
\bea
\text{Model DFT-A:}&&
(e^{2u}-H_{\Lambda}^2)(\hat{\mu}^2(G_0^2+G(u)^2))-\frac{1}{6}\sqrt{G_0}\lambda\Big((G_0^2+G(u)^2)\ln\Big(\frac{G_0^2+G(u)^2}{G_0^2}\Big)\nonumber\\
& &-1152\hat{\mu}^8e^{8u}(4u'(u'+1)+u'')\Big)=0,
\label{eq:DeFeliceModelAMena}
\\
\text{Model DFT-B:}&&(e^{2u}-H_{\Lambda}^2)(3\hat{\mu}^2(G_0^2+G(u)^2)^2)-G(u)^4\nonumber\\
&&+576\hat{\mu}^8e^{8u}G_0^2\Big((7u'-1)(u'+1)+2u''\Big)=0,
\label{eq:DeFeliceModelBMena}
\\
\text{Model DFT-C:}&&(e^{2u}-H_{\Lambda}^2)-\frac{\sqrt{G_0}\lambda}{3\hat{\mu}}\Big(\frac{-576\hat{\mu}^8e^{8u}(4u'(u'+1))+u'')}{G_0^2\cosh^2{\Big(\frac{G(u)}{G_0}}\Big)}\nonumber\\
&&+\frac{G(u)}{G_0}\tanh{\Big(\frac{G(u)}{G_0}\Big)}-\ln\Big[\cosh{\Big(\frac{G(u)}{G_0}\Big)}\Big]\Big)=0,
\label{eq:DeFeliceModelBMena}
\eea
where $H_{\Lambda}^2=(\tilde{\omega}_r e^{-4N} + \tilde{\omega}_m e^{-3N})+\frac{\Lambda}{3\hat{\mu}^2}$ and $G(u)=\hat{\mu}^4e^{4u}24(u'+1)$.

Unlike equations (\ref{eq:DeFeliceFreidmannA}), (\ref{eq:DeFeliceFreidmannB}), and (\ref{eq:DeFeliceFreidmannC}), we found that the equations written in terms of the logarithmic variables allow stable numerical integrations over redshift ranges from $z=5$ to $z=0$. As discussed in \cite{MoldenhauerIshak2009}, it is necessary to perform the integration forward in time (backward in the redshift) with initial conditions provided by some approximate solutions at high redshift. 
We verified that at higher redshifts ($z>5$), the approximate solutions provide an excellent fit to the numerical solution of the ODEs
with a relative difference $\frac{H_{approx}-H(z)}{H(z)}$ that is better than $0.1\%$. 
We use these approximations in order to find the numerical initial conditions for $u$ and $u'$ at $z=5$ and then start a numerical integration of the ODEs down to the supernovae locations. Proceeding in this way, we obtained very stable programs to derive various Hubble plots for the HOG models. The approximate solutions for the models of this section can be derived as 
\bea
\text{Model DFT-A:}&&H^2=\hat{\mu}^2H_{\Lambda}^2+\frac{\sqrt{G_0}\lambda}{6(G_0^2+G(\tilde{u})^2)}\Big((G_0^2+G(\tilde{u})^2)\ln\Big(\frac{G_0^2+G(\tilde{u})^2}{G_0^2}\Big)\nonumber\\
&&-1152\hat{\mu}^8e^{8\tilde{u}}(4\tilde{u}'(\tilde{u}'+1)+\tilde{u}'')\Big),
\label{eq:DeFeliceModelAMenaHapprox}
\\
\text{Model DFT-B:}&&H^2=\hat{\mu}^2H_{\Lambda}^2+\frac{G(\tilde{u})^4-576\hat{\mu}^8e^{8\tilde{u}}G_0^2\Big((7\tilde{u}'-1)(\tilde{u}'+1)+2\tilde{u}''\Big)}{3(G_0^2+G(\tilde{u})^2)^2},
\label{eq:DeFeliceModelBMenaHapprox}
\\
\text{Model DFT-C:}&&H^2=\hat{\mu}^2H_{\Lambda}^2+\frac{\sqrt{G_0}\lambda}{3}\Big(\frac{-576\hat{\mu}^8e^{8\tilde{u}}(4\tilde{u}'(\tilde{u}'+1))+\tilde{u}'')}{G_0^2\cosh^2{\Big(\frac{G(\tilde{u})}{G_0}}\Big)}\nonumber\\
&&+\frac{G(\tilde{u})}{G_0}\tanh{\Big(\frac{G(\tilde{u})}{G_0}\Big)}-\ln\Big[\cosh{\Big(\frac{G(\tilde{u})}{G_0}\Big)}\Big]\Big),
\label{eq:DeFeliceModelCMenaHapprox}
\eea
where  $G(\tilde{u})=\hat{\mu}^4e^{4\tilde{u}}24(\tilde{u}'+1)$  and the source $\tilde{u}$ is used from earlier.

The generalized Friedmann equations above along with their high redshift approximations are in a form ready for numerical integration and comparisons to observational data in \S 4.  
%
%
\subsection{$f(G)$ models proposed by Zhou, Copeland and Saffin}
%
The authors of \cite{ZhouCopeland2009} performed a thorough phase space analysis of $f(G)$ models and analyzed some specific models that satisfy some cosmological viability conditions. Following \cite{Amendola2006} for $f(R)$ models, the authors of \cite{ZhouCopeland2009} expressed the viability conditions as constraints on the derivatives of the function $f(G)$. We consider here some of their models as: 
\bea
\text{Model ZCS-A:}&&f(G)=\alpha \sqrt{G} + \beta \sqrt[4]{G},
\label{eq:ZhouCopelandModelA}
\\
\text{Model ZCS-B:}&&f(G)=\alpha (G^{3/4}-\beta)^{2/3},
\label{eq:ZhouCopelandModelB}
\\
\text{Model ZCS-C:}&&f(G) = \sqrt{\alpha} \exp (\beta/G).
\label{eq:ZhouCopelandExpG}
\eea
where $\alpha$ and $\beta$ are constants. The equations of motion in a flat FLRW spacetime follow as:

\noindent Model ZCS-A:
\bea
\lefteqn{3H^2=-\Big( \frac{1}{2}\alpha\sqrt{G}H(H^3-\frac{1}{2}(H'^2H+H''H^2))}\nonumber\\
& &+\frac{3}{4}\beta\sqrt[4]{G}H(\frac{1}{2}H'^2H+H^3+H^2H'-\frac{1}{4}(H'^2H+H''H^2))\Big)\Big/\Big(H'H+H^2\Big)\nonumber\\
&&+\frac{3H_0^2\Omega_m}{e^{3N}} + \frac{3H_0^2\Omega_{rad}}{e^{4N}}
\label{eq:ZhouCopelandModelAFriedmann}
\eea
Model ZCS-B:
\bea
\lefteqn{3H^2=\frac{1}{2}\alpha\Big(2(H^2(H'H+H^2))^{3/4}\beta^3 H(H'+H)-576H^{12}(H'+H)24^{1/4}}\nonumber\\
& &-\frac{1}{2}H^4(9H'^2+18H'H+10H^2)(H^2(H'H+H^2))^{1/2}24^{3/4}\nonumber\\
& & +12\beta H^7(5H'^3+15H+18H'H^2+8H^2)(H^2(H'H+H^2))^{1/4}24^{1/2}\nonumber\\
& & +\Big(288H^9(H'+H)^224^{1/4}+\frac{1}{4}H^3\beta^2(H^2(H'H+H^2))24^{3/4}\nonumber\\
&&-18H^6\beta(H'+H)(H^2(H'H+H^2))^{1/4}24^{1/2}\Big)(H'^2H+H''H^2)\Big)\Big/ \nonumber\\
&&\Big((H'H+H^2)(24^{3/4}(H^2(H'H+H^2))^{3/4}-\beta)^{7/3}(H^2(H'H+H^2))^{3/4}\Big) \nonumber\\
&& +\frac{3H_0^2\Omega_m}{e^{3N}} + \frac{3H_0^2\Omega_{rad}}{e^{4N}}
\label{eq:ZhouCopelandModelBFriedmann}
\eea
Model ZCS-C:
\bea
\lefteqn{3H^2=-\alpha \exp\Big(\frac{\beta }{24H^2/(H'H+H^2)}\Big)\Big((288H^{10}(H'+H)^2+2\beta^2H^2H'(H'+2H)}\nonumber\\
& &+(H'^2H+H''H^2)(24H^4\beta(H'+H)+\beta^2H-144H^7(H'+H)^2))(24H^3H'+24H^4)^{1/2}\nonumber\\
& &  +24H^5\beta(H^3+9H'^2H+7H'H^2+3H'^3)\Big) \Big/\Big(576H^4(H'H+H^2)^4\Big)\nonumber\\
&&+\frac{3H_0^2\Omega_m}{e^{3N}} + \frac{3H_0^2\Omega_{rad}}{e^{4N}}
\label{eq:ZhouCopelandEXPGBFriedmann}
\eea
Again, in order to compare these models to cosmological data, we must write the generalized Freidmann equation in a numerically stable integrable form using logarithmic variables yielding: 

\noindent Model ZCS-A:
\bea
(e^{2u}&-&e^{2\tilde{u}})(48\hat{\mu}^2(u'+1)^2)-4\alpha \sqrt{G(u)}(u''+2(u'+1)(u'-1))\nonumber\\
&&-3\beta \sqrt[4]{G(u)}(u''-4(u'+1))=0,
\label{eq:ZhouCopelandModelAMena}
\eea
Model ZCS-B:
\bea
\lefteqn{(e^{2u}-e^{2\tilde{u}})6\hat{\mu}^2(u'+1)(G(u)(G(u)^{3/4}-\beta)^{7/3}) -(G(u)^{13/4}(u'-1)}\nonumber\\ 
& &+G(u)^{5/2}\beta (u'+4)-G(u)^{7/4}\beta^2(4u'+5)+2G(u)\beta^3(u'+1))\nonumber\\
&&-6\hat{\mu}^4e^{4u}(2G(u)^{9/4}-3G(u)^{3/2}\beta +G(u)^{3/4}\beta^2)u''=0,
\label{eq:ZhouCopelandModelBMena}
\eea
Model ZCS-C:
\bea
\lefteqn{(e^{2u}-e^{2\tilde{u}})(3\hat{\mu}^2(u'+1)^2G(u)^2) + \alpha \sqrt{G(u)} \exp(\beta /G(u))\Big(\Big(-\frac{1}{2}G(u)^2(u'-1)}\nonumber\\
&&+G(u)\beta (5u'+1) +4\beta^2u'\Big)(u'+1)+\Big(\beta^2 - \frac{1}{4}G(u)^2+G(u)\beta\Big)u''\Big) =0,
\label{eq:ZhouCopelandExpGBMena}
\eea
We also need the approximate solution for these models at high redshift providing initial conditions for numerical integrations as discussed in previous section. The approximate solutions at high redshift for these models are:

\noindent Model ZCS-A:
\bea
H^2 = \hat{\mu}^2e^{2\tilde{u}}+\frac{4\alpha \sqrt{G(\tilde{u})}(\tilde{u}''+2(\tilde{u}'+1)(\tilde{u}'-1))+3\beta \sqrt[4]{G(\tilde{u})}(\tilde{u}''-4(\tilde{u}'+1))}{48(\tilde{u}'+1)^2},
\label{eq:ZhouCopelandModelAHapprox}
\eea
Model ZCS-B:
\bea
H^2&=& \hat{\mu}^2e^{2\tilde{u}}+\Big((G(\tilde{u})^{13/4}(\tilde{u}'-1)+G(\tilde{u})^{5/2}\beta (\tilde{u}'+4)-G(\tilde{u})^{7/4}\beta^2(4\tilde{u}'+5)\nonumber\\
&&+2G(\tilde{u})\beta^3(\tilde{u}'+1)) +6\hat{\mu}^4e^{4\tilde{u}}(2G(\tilde{u})^{9/4}-3G(\tilde{u})^{3/2}\beta \nonumber\\
& & +G(\tilde{u})^{3/4}\beta^2)\tilde{u}''\Big) \Big/ \Big(6\hat{\mu}(G(\tilde{u})^{3/4}-\beta)^{7/3}G(\tilde{u})(\tilde{u}'+1)\Big),
\label{eq:ZhouCopelandModelBHapprox}
\eea
Model ZCS-C:
\bea
H^2&=& \hat{\mu}^2e^{2\tilde{u}}+\alpha \sqrt{G(\tilde{u})} \exp(\beta /G(\tilde{u}))\Big(\Big(-\frac{1}{2}G(\tilde{u})^2(\tilde{u}'-1)+G(\tilde{u})\beta (5\tilde{u}'+1) +4\beta^2\tilde{u}'\Big)(\tilde{u}'+1)\nonumber\\
& &+\Big(\beta^2 - \frac{1}{4}G(\tilde{u})^2+G(\tilde{u})\beta\Big)\tilde{u}''\Big) \Big/ \Big(3\hat{\mu}^2(\tilde{u}'+1)^2G(\tilde{u})^2\Big),
\label{eq:ZhouCopelandExpGBHapprox}
\eea
where again $G(\tilde{u})=\hat{\mu}^4e^{4\tilde{u}}24(\tilde{u}'+1)$ and the source $\tilde{u}$ is as defined earlier.
%
\subsection{$f(G)$ Models proposed by Uddin, Lidsey and Tavakol }
%
The models presented by \cite{Uddin2009} are similar to the models presented by \cite{ZhouCopeland2009} of the previous section, i.e. $f(G)=\alpha \sqrt{G}+\beta \sqrt[4]{G}$, with $\beta=0$ and $\alpha=\sqrt{4\alpha}$, although \cite{Uddin2009} performed a different and independent analysis. In agreement with the analogous scalar field models, the power-law form of the $f(G)$ model as $f(G)=\pm 2\sqrt{\alpha G}$ was found to have stable scaling solutions.  The authors studied the equation of state parameter for these models as it evolved through radiation, matter and accelerating epochs.  The behavior of the energy densities were also discussed there. By investigating stability for both vacuum and non-vacuum solutions,  it was recognized that those models suffered from a singular point at transition. Our results for these models can be found with results for the ZCS models in \S 5 below.     
\begin{figure}
\begin{center}
\begin{tabular}{|c|c|}
\hline
{\includegraphics[width=2.4in,height=3.2in,angle=-90]{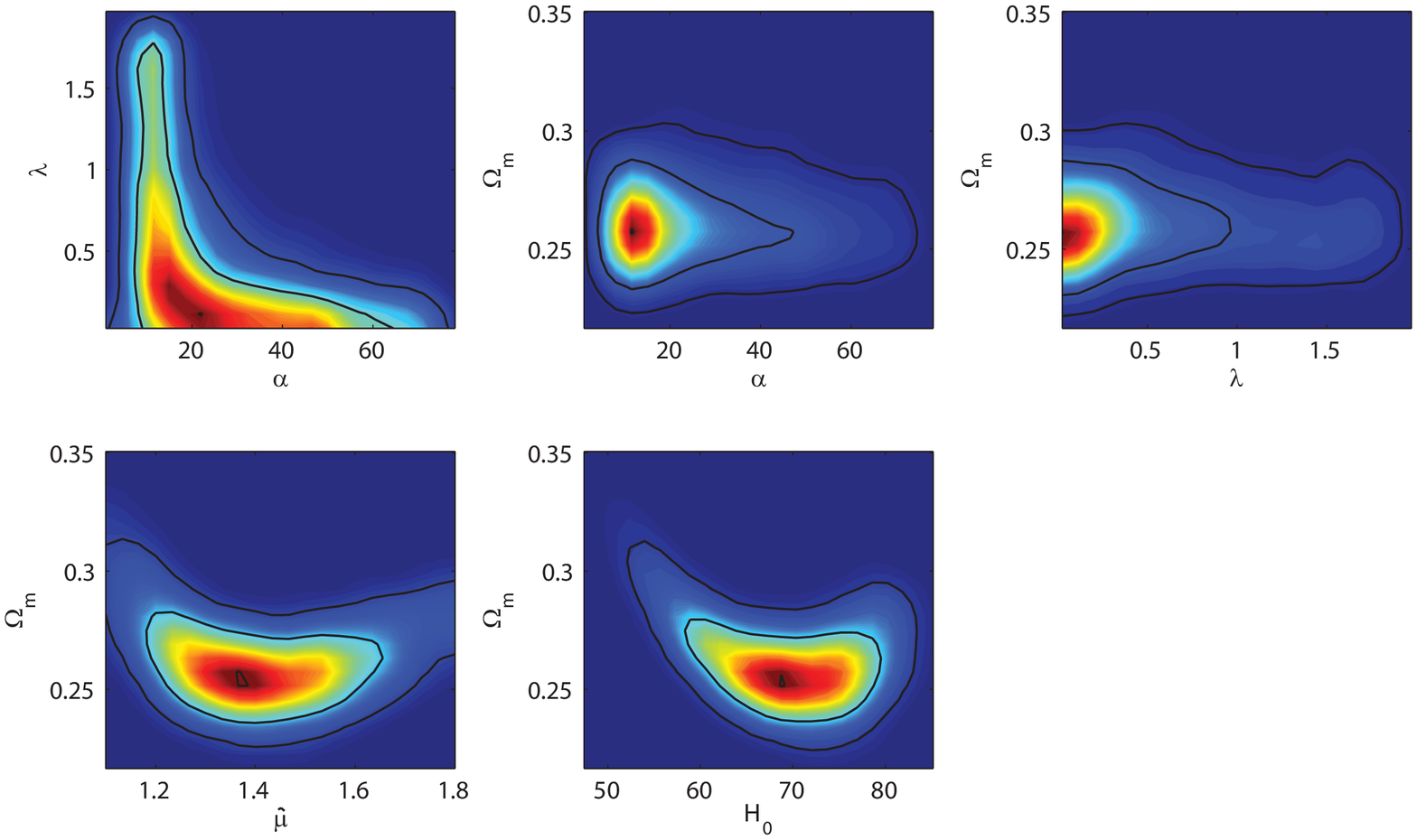}} &
{\includegraphics[width=2.4in,height=3.2in,angle=-90]{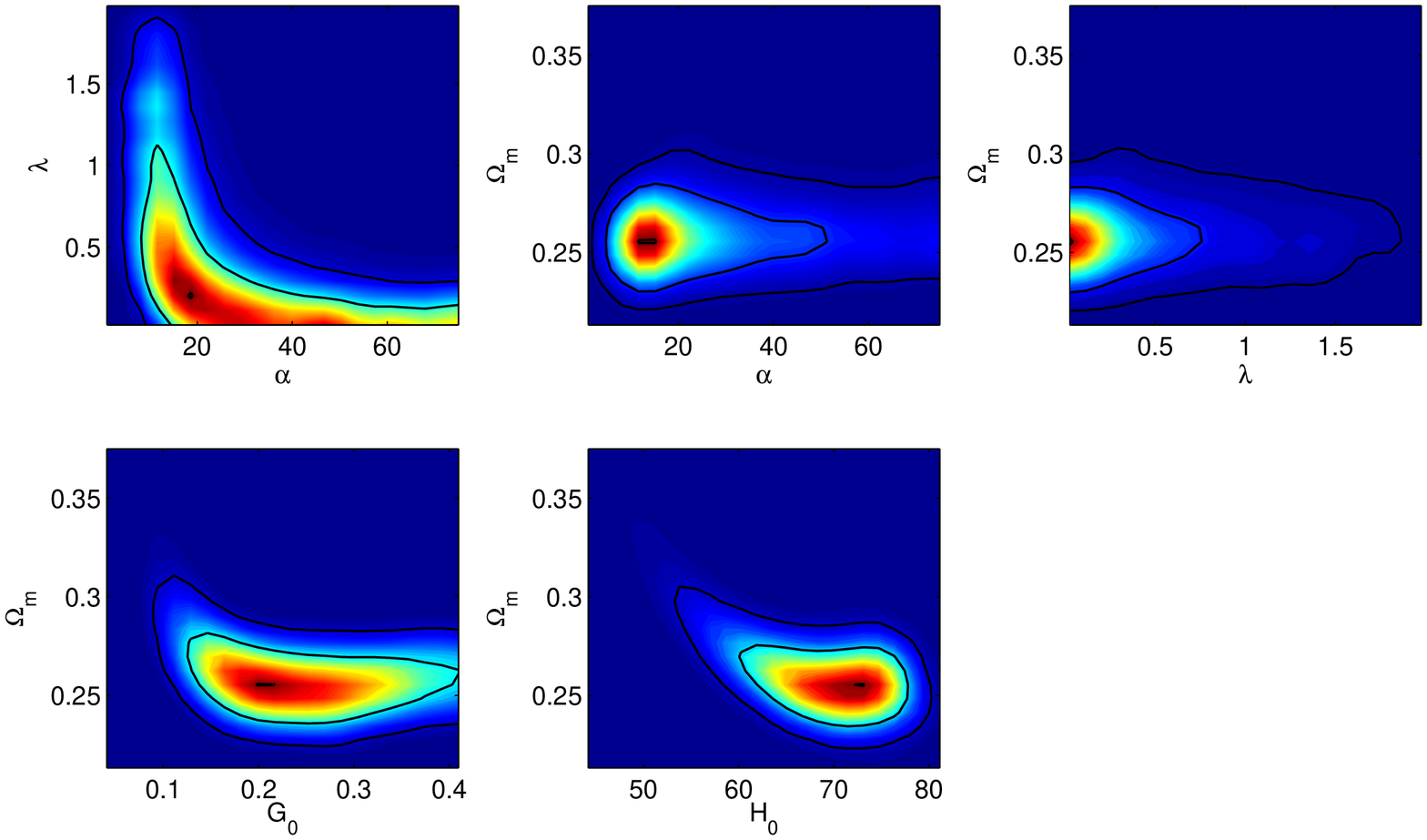}} \\
\hline
\end{tabular}
\caption{
LEFT: $2D$ joint contour plots for DFT-A with parameters $\Omega_m$, $\alpha$, $\lambda$, $\hat{\mu}$, $H_0$ for Union, WMAP5 and SDSS LRG(BAO) data sets, where the inner and outer loops are $68\%$ and $95\%$, respectively.  RIGHT:$2D$ joint contour plots for DFT-A with parameters $\Omega_m$, $\alpha$, $\lambda$, $G_0$, $H_0$ for Union, WMAP5 and SDSS LRG(BAO) data sets, where the inner and outer loops are $68\%$ and $95\%$, respectively.} 
\end{center}
\end{figure}
%
%
%
\section{Cosmological distances constraints from SNe Ia, CMB surface, and BAO}
%
In this section, we describe the three cosmological observations used to constrain the models described above. One of the first compelling evidences for cosmic acceleration came from the Supernovae type Ia (SNe Ia) observations and we use here the distance modulus as a function of the redshift $z$ given by 
\be
\mu(z)=\tilde{m}-M=5\log{D_L}+25
\label{eq:DistanceModulus}
\ee
containing the magnitude-redshift function, $\tilde{m}(z)$ and a nuisance parameter, $M$ which is degenerate with the Hubble parameter, $H_0$. The luminosity distance, $D_L$ in units of Mpc for HOG models is given by
\be
D_L(z)=(1+z)\int^z_0 \frac{1}{H(z')}dz'\,,
\label{eq:LuminosityDistance}
\ee
with $H(z')$ as the solution to the non-linear differential equations given above for the higher order gravity models (recall that for the logarithmic variables defined in the previous section the redshift reads as $z=e^{-N}-1$.)  
We use the Union set of supernovae which was compiled as an attempt to gather the best of the best supernovae from different surveys, including Supernovae Legacy Survey, ESSENCE Survey, HST, and other older sets \cite{Kowalski2008}.  After selection cuts, the 414 SNe Ia are reduced to 307. The fitting of the SNe Ia uses a standard minimization method, $\chi^2$,  
\be
\chi_{SN}^2=\sum^{i=307}_{i=1} \frac{[\mu^i_{HOG}(z)-\mu^i_{obs}(z)]^2}{\sigma^2}
\label{eq:ChiSquared}
\ee
where $\sigma$ is the magnitude uncertainty and $i$ the number of data points compared. We break degeneracies in the parameter space by using the value of the Hubble parameter as measured by the Hubble Key Project (HST)\cite{HubbleKeyProject}, $H_0=72\pm8 \,km/s/Mpc$.

Next, we consider the distance to the CMB last scattering surface and, following \cite{Komatsu2008}, we define three fitting parameters for comparison to the WMAP5 data using the shift parameter, $R$, see for example \cite{Bond}, 
\be
R(z_*)=\sqrt{\Omega_m}H_0(1+z_*)D_A(z_*),
\label{eq:ShiftParameter}
\ee
with the redshift, $z_*$ for the surface of last scattering, see for example \cite{HuSugiyama},
\be
z_*=1048[1+0.00124(\Omega_b h^2)^{-0.738}][1+g_1 (\Omega_m h^2)^{g_2}],
\label{eq:zstar}
\ee
where 
\be
g_1=\frac{0.0783(\Omega_b h^2)^{-0.238}}{1+39.5(\Omega_b h^2)^{0.763}},
\label{eq:zstarg1}
\ee
and
\be
g_2=\frac{0.560}{1+21.1(\Omega_b h^2)^{1.81}},
\label{eq:ztarg2}
\ee
and third, the acoustic scale, $l_a$, is, see for example \cite{WangMukherjee, Wright},
\be
l_a=(1+z_*)\frac{\pi D_A(z_*)}{r_s(z_*)},
\label{eq:AcousticScale}
\ee
with the proper angular diameter distance, $D_A(z)=D_L(z)/(1+z)^2$ and the comoving sound horizon, $r_s(z_*)$, see for example \cite{Percival2007}, 
\be
r_s(z_*)=\frac{1}{\sqrt{3}}\int^{1/(1+z_*)}_{0}{\frac{da}{a^2H(a)\sqrt{1+(3\Omega_b/4\Omega_{\gamma})a}}},
\label{eq:SoundHorizon}
\ee
with $\Omega_{\gamma}=2.469\times 10^{-5}h^{-2}$ for $T_{cmb}=2.725 K$.
\begin{figure}
\begin{center}
\begin{tabular}{|c|c|}
\hline

{\includegraphics[width=2.2in,height=3.1in,angle=-90]{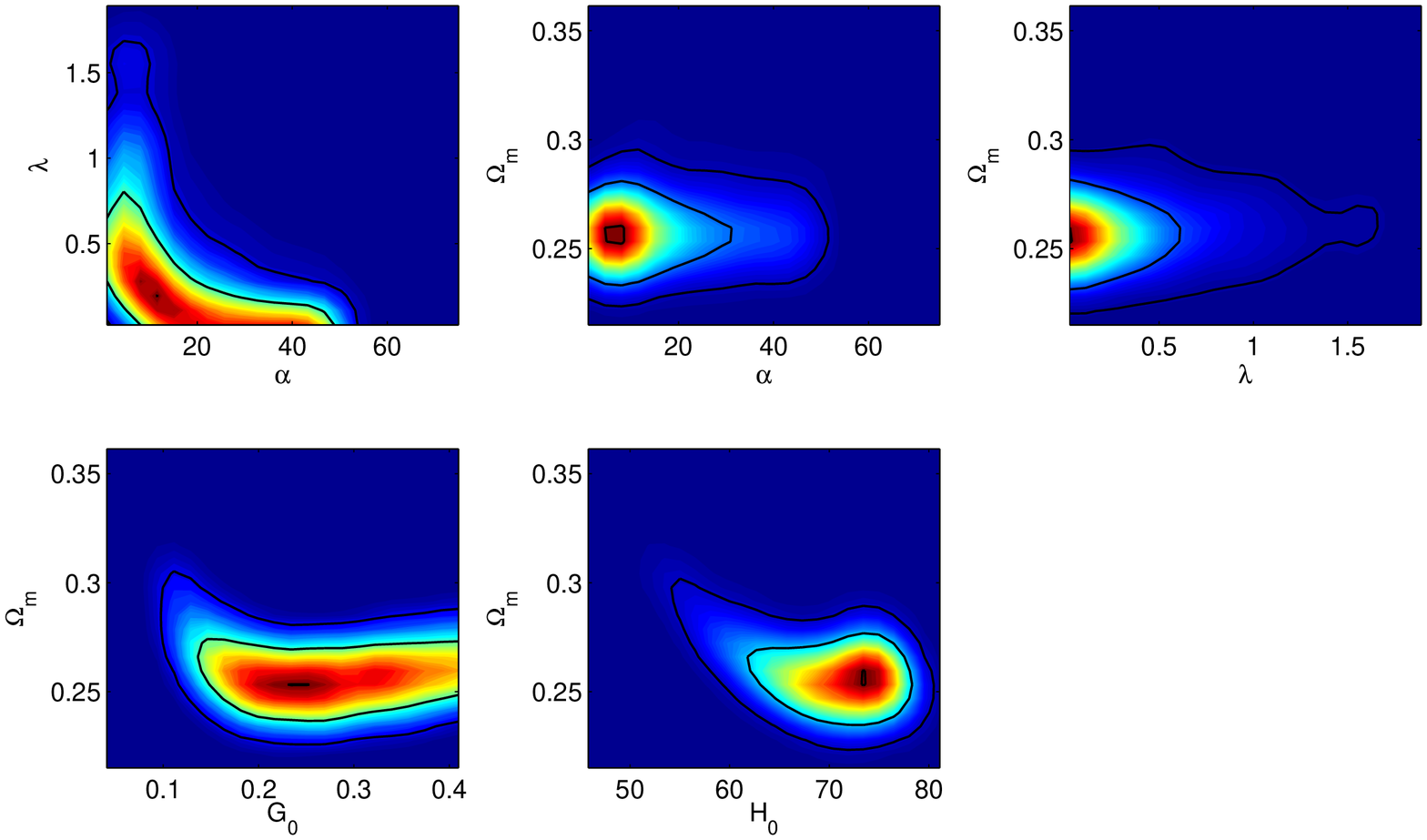}} &
{\includegraphics[width=2.2in,height=3.1in,angle=-90]{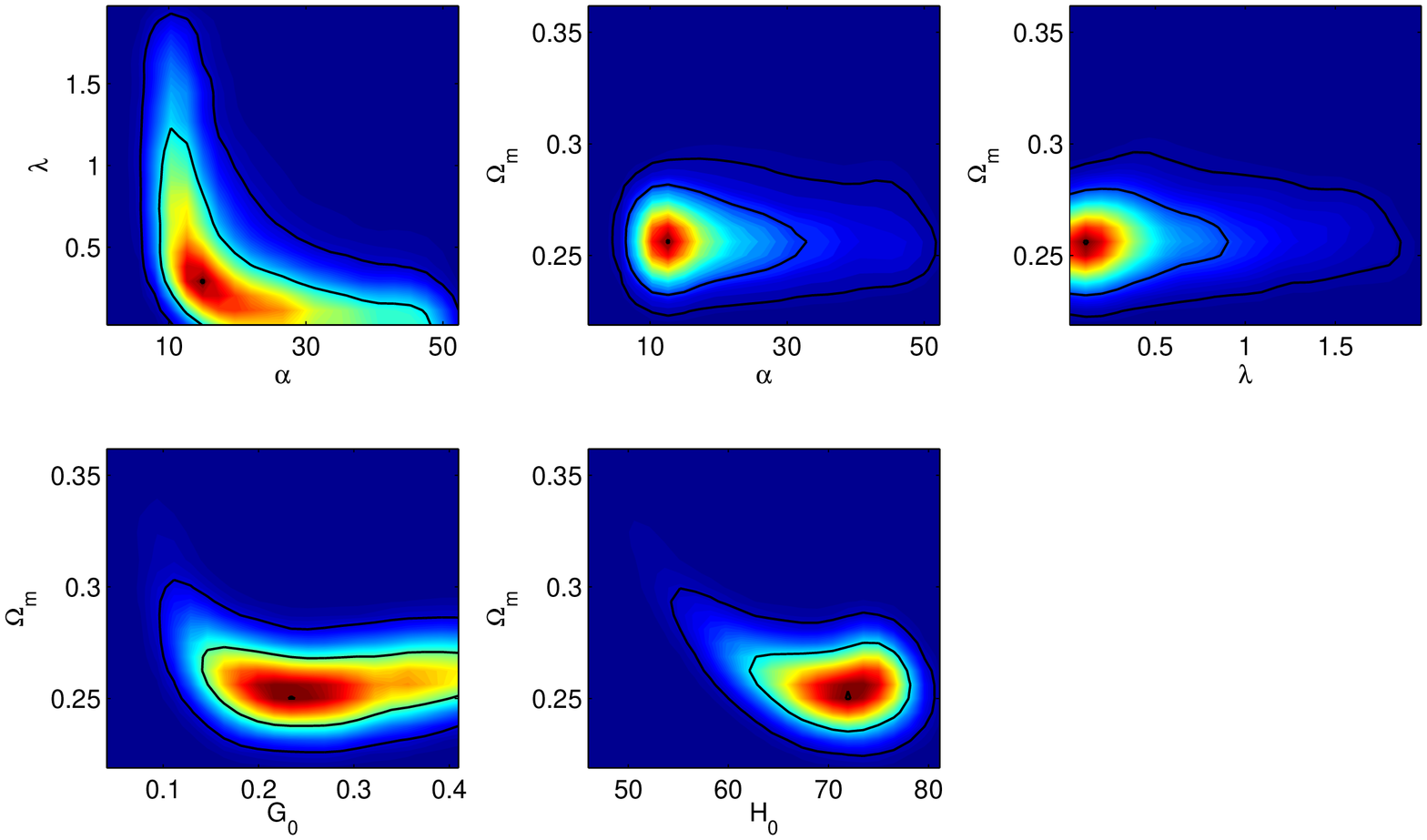}} \\
\hline
\end{tabular}
\caption{
LEFT: $2D$ joint contour plots for DFT-B with parameters $\Omega_m$, $\alpha$, $\lambda$, $G_0$, $H_0$ for Union, WMAP5 and SDSS LRG(BAO) data sets, where the inner and outer loops are $68\%$ and $95\%$, respectively.  RIGHT:$2D$ joint contour plots for DFT-C with parameters $\Omega_m$, $\alpha$, $\lambda$, $G_0$, $H_0$ for Union, WMAP5 and SDSS LRG(BAO) data sets, where the inner and outer loops are $68\%$ and $95\%$, respectively.} 
\end{center}
\end{figure}
Together the parameters $x_i=(R,l_a,z_*)$ are used to fit $\chi^2_{WMAP}=\triangle x_iCov^{-1}(x_ix_j)\triangle x_j$ with $\triangle x_i=x_i-x^{obs}_i$ and $Cov^{-1}(x_ix_j)$ is the inverse covariance matrix for the parameters. 

Thirdly, we use BAO constraints following \cite{Komatsu2008, Percival2007, Eisenstein2005, Tegmark} and defining the ratio of the sound horizon, $r_s(z_d)$ to the effective distance, $D_V$ as a fit for SDSS by
\be
\chi^2_{BAO}=\Big(\frac{r_s(z_d)/D_V(z=0.2)-0.198}{0.0058}\Big)^2+\Big(\frac{r_s(z_d)/D_V(z=0.35)-0.1094}{0.0033}\Big)^2,
\label{eq:chisqbao}
\ee
with, see for example \cite{Eisenstein2005}, 
\be
D_V(z)=\Big(D_A^2(z)(1+z)^2\frac{z}{H(z)}\Big)^{1/3},
\label{eq:EffectiveDistance}
\ee
and the redshift, $z_d$ as, see for example \cite{EisensteinHu1998},
\be
z_d=\frac{1291(\Omega_m h^2)^{0.251}}{1+0.659(\Omega_m h^2)^{0.828}}[1+b_1(\Omega_b h^2)^{b_2}],
\label{eq:zdrag}
\ee
where
\be
b_1=0.313(\Omega_m h^2)^{-0.419}[1+0.607(\Omega_m h^2)^{0.674}],
\label{eq:zdragb1}
\ee
and
\be
b_2=0.238(\Omega_m h^2)^{0.223}.
\label{eq:zdragb2}
\ee

We use a modified and extended version of the publicly 
available package, CosmoMC \cite{cosmomc} to perform a Monte Carlo Markov Chain analysis (MCMC's). The MCMC's are used to compute the likelihoods for 
the parameters in the model.  This method randomly chooses values for the above parameters and, based on the $\chi^2$ obtained, either accepts or rejects the set of parameters via the Metropolis-Hastings algorithm.  When a set of parameters is accepted it is added to the chain and forms a new starting point for the next step.  The process is repeated until the specified convergence is reached. 
We found that an elaborate and essential step in our analysis was to numerically integrate the stiff ODEs in order to derive the Hubble expansion rate for a wide range of redshift and integrate it to the CosmoMC package. For this task, we made a necessary change of variable in our integration and also used a good approximation at higher redshifts in order to obtain initial values for our codes to perform stable integrations from $z=5$ down to $z=0$. 
We used the framework developed to compare the models described in the previous sections to the three sets of observations and the results are given in the next section. 
\begin{figure}
\begin{center}
\begin{tabular}{|c|c|c|}
\hline

{\includegraphics[width=2.0in,height=2.4in,angle=0]{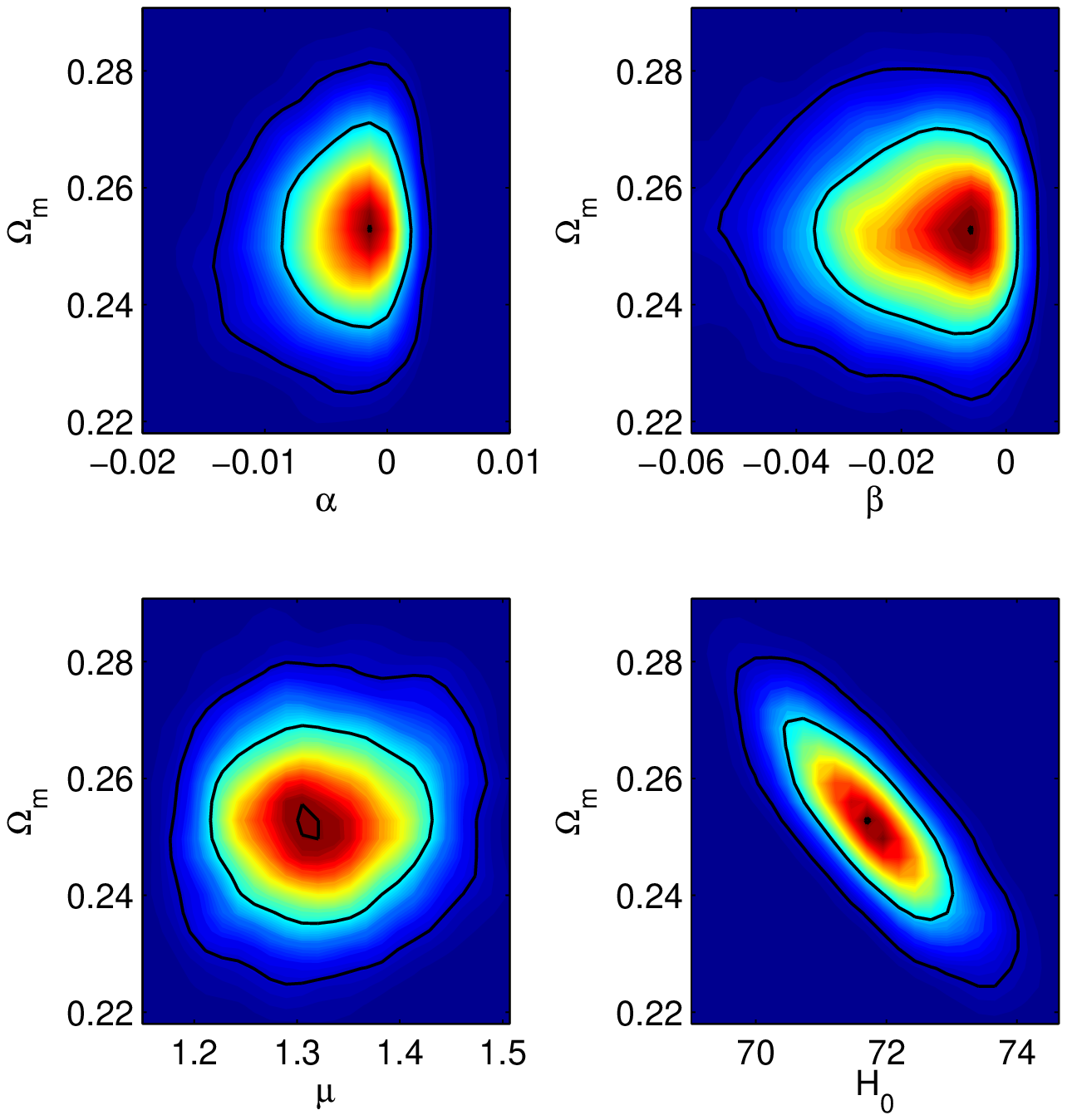}} &
{\includegraphics[width=2.0in,height=2.4in,angle=0]{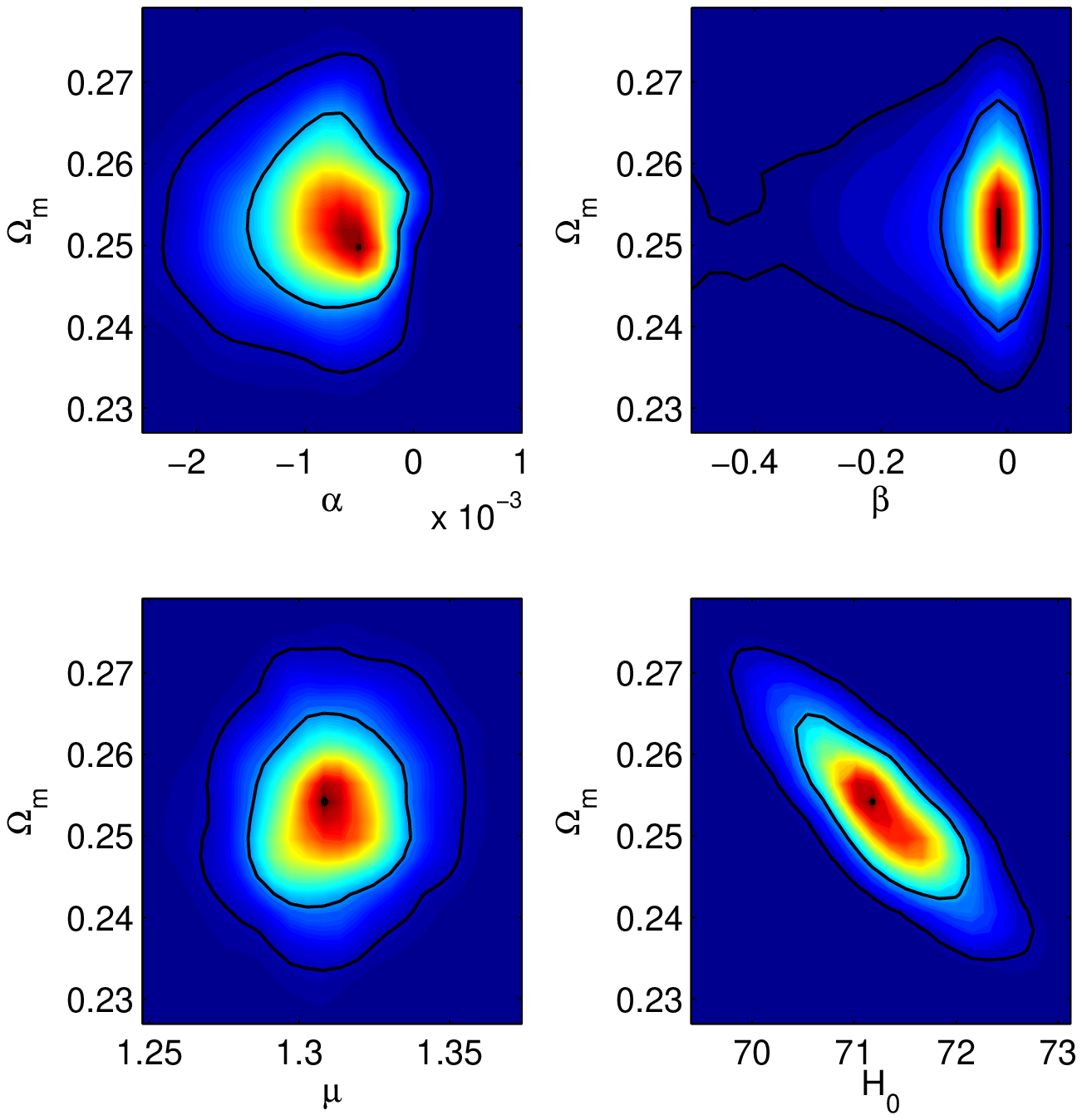}} &
{\includegraphics[width=2.0in,height=2.4in,angle=0]{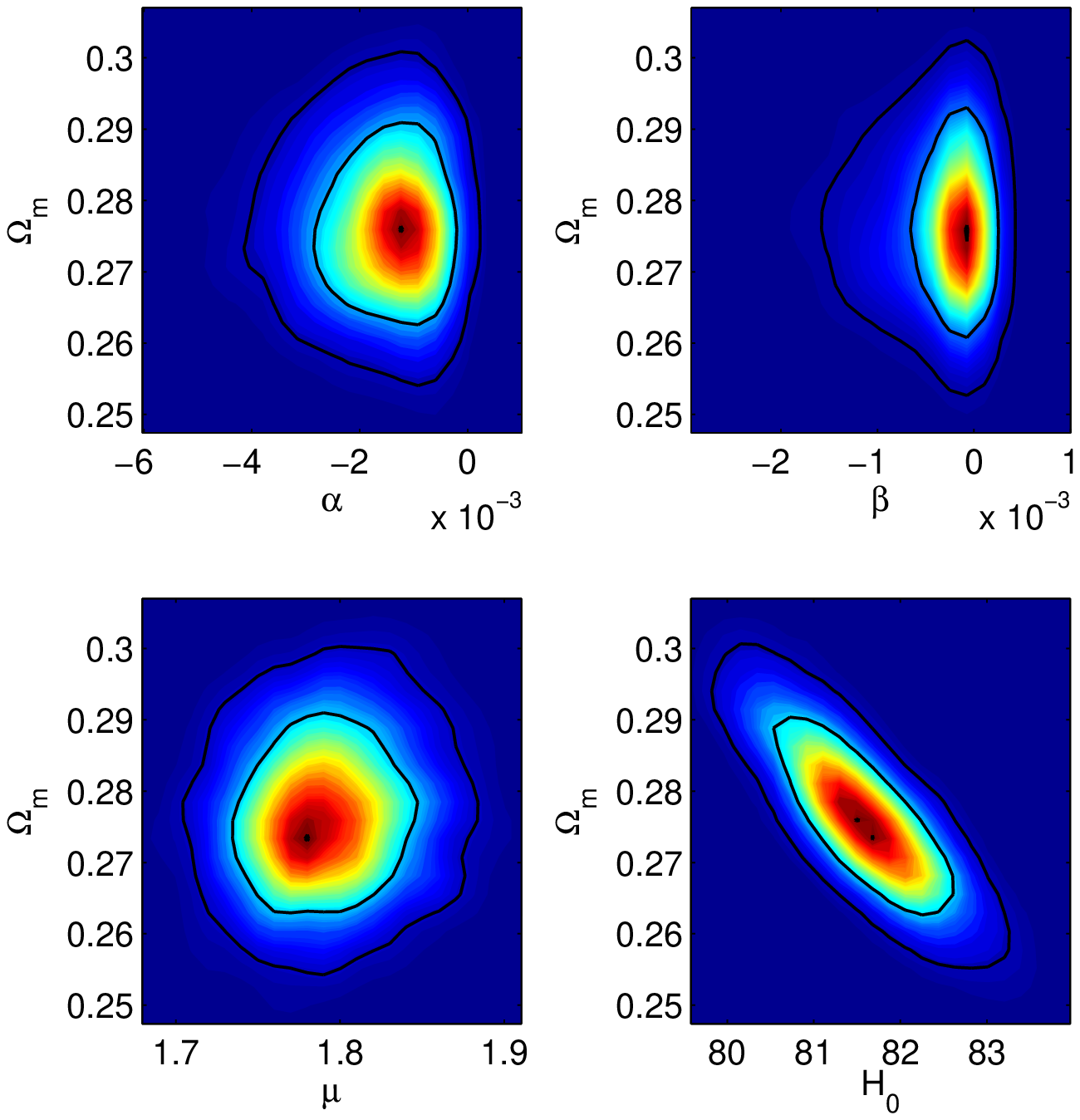}} \\
\hline
\end{tabular}
\caption{
LEFT: $2D$ joint contour plots for ZCS-A with parameters $\Omega_m$, $\alpha$, $\beta$, $H_0$ for Union, WMAP5 and SDSS LRG(BAO) data sets, where the inner and outer loops are $68\%$ and $95\%$, respectively. CENTER:$2D$ joint contour plots for ZCS-B with parameters $\Omega_m$, $\alpha$, $\beta$, $H_0$ for Union, WMAP5 and SDSS LRG(BAO) data sets, where the inner and outer loops are $68\%$ and $95\%$, respectively. RIGHT:$2D$ joint contour plots for ZCS-C with parameters $\Omega_m$, $\alpha$, $\beta$, $H_0$ for Union, WMAP5 and SDSS LRG(BAO) data sets, where the inner and outer loops are $68\%$ and $95\%$, respectively.} 
\end{center}
\end{figure}
%
%

\section{Results and Conclusion}
%
%
\begin{figure}
\begin{center}
\begin{tabular}{|c|c|}
\hline
{\includegraphics[width=2.8in,height=2.8in,angle=-90.]{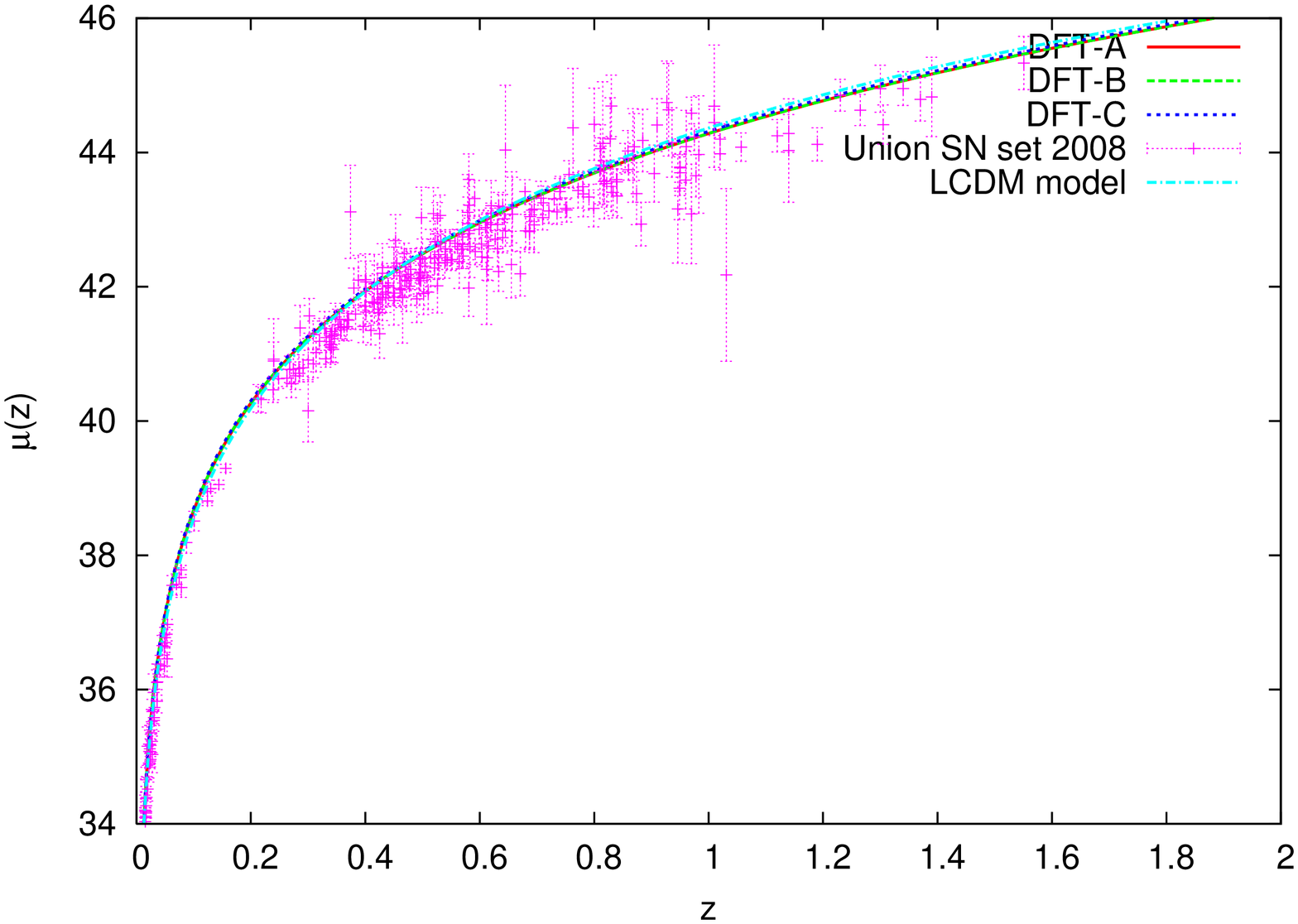}}  &
{\includegraphics[width=2.8in,height=2.8in,angle=-90.]{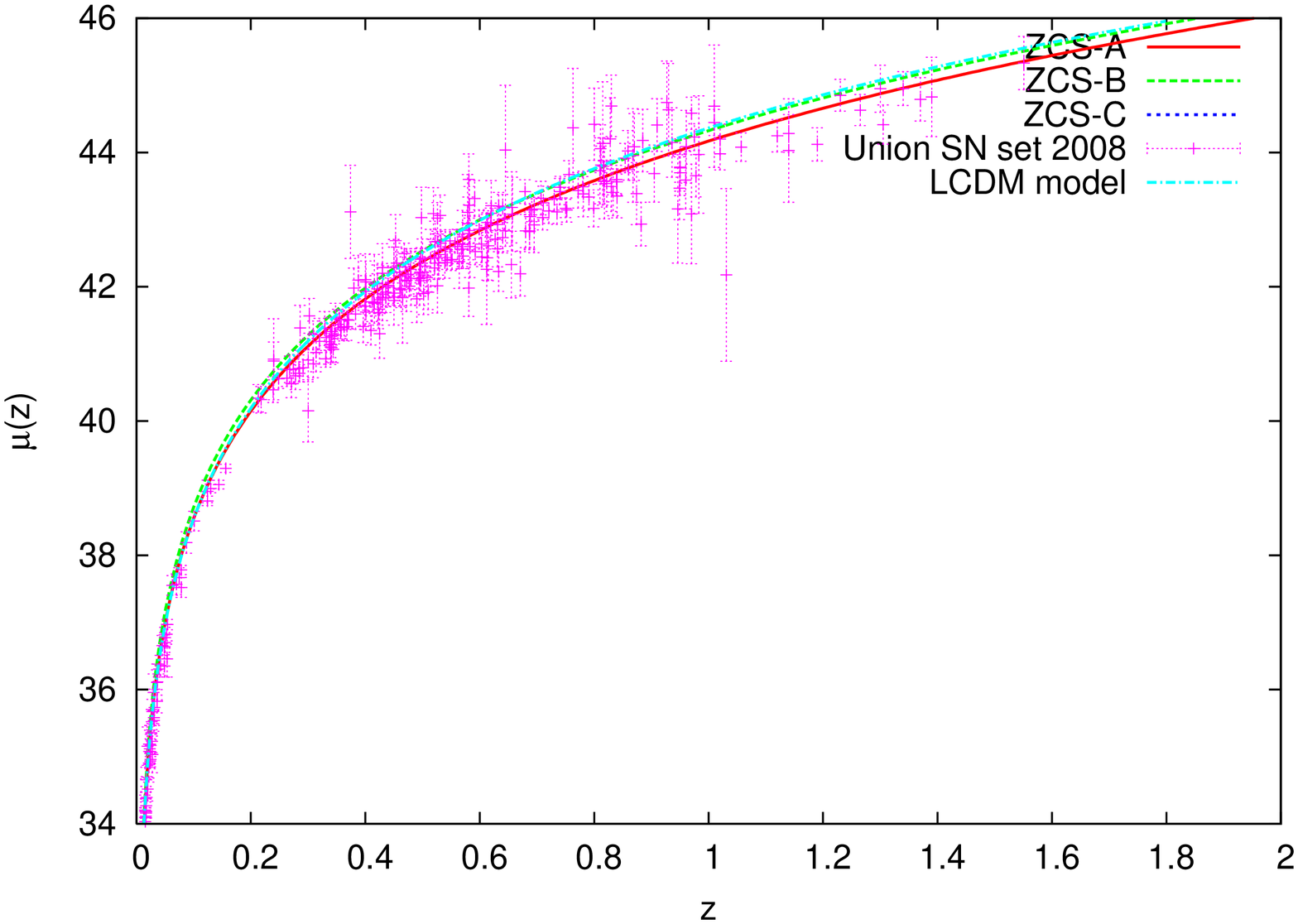}} \\
\hline
\end{tabular}
\caption{
Comparisons of different SNe Ia plots with Union data and different values of $f(G)$ models. LEFT: Comparison of DFT-A, DFT-B, DFT-C Models and LCDM. RIGHT:Comparison of ZCS-A, ZCS-B, ZCS-C Models and LCDM.} 
\end{center}
\end{figure}
Our results are summarized in figures 1-4 and in table I. In figure 1, we present the results for model DFT-A. We first (left part of figure 1) varied  $\Omega_m$, $\alpha$, $\lambda$, $\hat{\mu}$, and $H_0$ leaving $G_0\sim h^4$ as a derived parameter, and we obtained a normalized $\chi^2/dof = 1.0754$ ($\chi^2=329.1$). On the right part of figure 1 for model DFT-A, we varied $\Omega_m$, $\alpha$, $\lambda$, $G_0$, and $H_0$ leaving $\hat{\mu}$ as a derived parameter, with a normalized $\chi^2/dof = 1.0751$ ($\chi^2=329.0$). Our best fit values are given in table 1 and we find an $\Omega_m$ around $0.25$ in both cases. We find best values for the model parameters $\alpha$ and $\lambda$ that have a product $\alpha\lambda$ of the order unity as required, see \cite{DeFelice2008}. We find the best fit for $\hat{\mu}$ that is of the order of twice the normalized Hubble parameter $h$, as it should \cite{Mena2006,MoldenhauerIshak2009}, and the best fit parameter $G_0$ of the order of $h^4$ \cite{DeFelice2008}.  Next, we show in figure 2 the results for models DFT-B and DFT-C with similar results and normalized $\chi^2/dof = 1.0764$ ($\chi^2=329.4$) and $\chi^2/dof = 1.0748$ ($\chi^2=328.9$) respectively. Finally, we show in figure 3, results for the ZCS models varying the parameters $\alpha$, $\beta$, $\hat{\mu}$, $\Omega_m$ and $H_0$. 
\begin{table*}[t]   
\begin{center}

\resizebox{6.4in}{!} {
\begin{tabular}{|c|c|c|c|c|c|c|c|c|c|}
\hline
 Constraints from&$\chi^2$&$\chi^2/dof$&$\Omega_m$ & $H_0$&$\hat{\mu}$ & $G_0$ &$\alpha$&$\lambda$&$\beta$ \\Observations & & &  &&&&&& \\
\hline
\hline 

$DFT-A$
                    &$329.1$&$1.0754$&$0.254939^{+0.049835}_{-0.025707}$&$71.2827^{+11.8991}_{-17.8981}$&$1.4424^{+0.356636}_{-0.309619}$&$-$&$26.6096^{+48.3636}_{-21.5027}$&$0.136131^{+1.75928}_{-0.111111}$& $-$\\
\hline
$DFT-A$
                    &$329.0$&$1.0751$&$0.253516^{+0.0607013}_{-0.0231282}$&$68.1122^{+11.3855}_{-16.2845}$&$-$&$.214908^{+0.194690}_{-0.114864}$&$54.3437^{+20.6530}_{-44.9701}$&$0.0832318^{+1.81255}_{-0.058080}$& $-$\\
\hline
$DFT-B$
                    &$329.4$&$1.0764$&$0.251225^{+0.0537648}_{-0.0218609}$&$70.6107^{+8.94104}_{-17.1149}$&$-$&$.253771^{+0.155822}_{-0.149289}$&$9.99891^{+39.9510}_{-6.78136}$&$0.313781^{+1.49185}_{-0.288712}$&$-$\\ 
\hline
$DFT-C$
                    &$328.9$&$1.0748$&$0.253603^{+0.049532}_{-0.023303}$&$69.6979^{+9.64599}_{-16.0135}$&$-$&$.238717^{+0.170843}_{-0.135459}$&$13.9569^{+36.0160}_{-4.56533}$&$0.529189^{+1.36409}_{-0.473996}$&$-$\\ 
\hline
$ZCS-A$
                    &$328.2$&$1.0725$&$0.25047^{+0.02857}_{-0.02093}$&$72.066^{+1.6351}_{-2.1441}$&$1.35273^{+0.13186}_{-0.18860}$&$-$&$0.00084^{+0.00016}_{-0.01632}$&$-$&$-0.03498^{+0.03595}_{-0.02399}$\\ 
\hline
$ZCS-B$
                    &$362.4$&$1.1843$&$0.254855^{+0.018011}_{-0.025755}$&$71.1289^{+1.81457}_{-1.37513}$&$1.30822^{+0.0448309}_{-0.0363830}$&$-$&$-0.00014^{+0.000083}_{-0.001439}$&$-$&$0.00031^{+0.03288}_{-0.76993}$\\ 
\hline
$ZCS-C$
                    &$332.4$&$1.0862$&$0.27302^{+0.02810}_{-0.01934}$&$81.511^{+1.8530}_{-1.7468}$&$1.7754^{+0.10141}_{-0.07048}$&$-$&$-0.00012^{+0.000004}_{-0.00295}$&$-$&$-0.0000017^{+0.00012}_{-0.00142}$\\ 
\hline
\end{tabular}
 }
 \caption{$\chi^2$, $\chi^2/dof$, and best-fit parameters for $f(G)$ higher order gravity models using observational constraints from supernovae, HST, CMB surface and BAO.  For models DFT-A, DFT-B, DFT-C the parameters are $\Omega_m$, $H_0$, $\hat{\mu}$, $G_0$, $\alpha$, and $\lambda$. For models ZCS-A, ZCS-B, ZCS-C the parameters are $\Omega_m$, $H_0$, $\hat{\mu}$, $\alpha$, and $\beta$.  }
\end{center}
\end{table*}

The best fit parameters are given in table 1. The first and second models have best fit values similar to the DFT models while the third model has slightly higher values for $\Omega_m$ and $H_0$. The best fit $\chi^2$'s for $f(G)$ models are close to the $\chi_{SN+BAO+CMB}^2\approx314.6$ ($\chi^2/dof=1.018$) that we obtained for the LCDM concordance model. In view of the of the possible systematic uncertainties in the supernova data, it is not clear that the difference between the $\chi^2$s is significant. It is worth noting that the parameter space that we found for the DFT models is also all contained within the parameter space found by \cite{DeFeliceSolar} for the models to be  compatible with solar system constraints. In other words, the DFT models analyzed in this paper pass physical acceptability conditions \cite{DeFelice2008, DeFelice2009}, solar system tests \cite{DeFeliceSolar} and here pass constraints from supernova, BAO rulers and distance to CMB last scattering surface. Recently, Ref. \cite{DeFeliceMota2009} pointed out in a preliminary work that matter perturbations in Gauss-Bonnet models exhibit some instabilities during the matter era, and that for the growth to be compatible with observations the  deviations from general relativity have to be very small. This point needs further investigations using perturbation studies in $f(G)$ models. In view of the success of $f(G)$ models with solar system tests and cosmological distances constraints, we conclude that these models need to be subjected, in future projects, to full large scale structure constraints such as galaxy clustering and gravitational lensing, as well as the full CMB analysis.  
\acknowledgments 
%
The authors thank B. Troup and J. Scott for useful discussions about the CosmoMC package. MI acknowledges that this material is based upon work supported in part by NASA under grant NNX09AJ55G and that part of the calculations for this work have been performed on the Cosmology Computer Cluster funded by the Hoblitzelle Foundation.  DE is supported in part by the World Premier International Research 
Center Initiative (WPI Initiative), MEXT, Japan and by a Grant-in-Aid for Scientific Research 
(21740167) from the Japan Society for Promotion of Science (JSPS), and by funds from the Arizona State University Foundation.
%

\end{document}